# Muppet: MapReduce-Style Processing of Fast Data

Wang Lam[1], Lu Liu[1], STS Prasad[1], Anand Rajaraman[1], Zoheb Vacheri[1], AnHai Doan[1,2]
{wlam,luliu,stsprasad,zoheb,anhai}@walmartlabs.com and anand@anandr.com
[1]@WalmartLabs, [2]University of Wisconsin-Madison

## ABSTRACT

MapReduce has emerged as a popular method to process big data. In the past few years, however, not just big data, but fast data has also exploded in volume and availability. Examples of such data include sensor data streams, the Twitter Firehose, and Facebook updates. Numerous applications must process fast data. Can we provide a MapReduce-style framework so that developers can quickly write such applications and execute them over a cluster of machines, to achieve low latency and high scalability?

In this paper we report on our investigation of this question, as carried out at Kosmix and WalmartLabs. We describe MapUpdate, a framework like MapReduce, but specifically developed for fast data. We describe Muppet, our implementation of MapUpdate. Throughout the description we highlight the key challenges, argue why MapReduce is not well suited to address them, and briefly describe our current solutions. Finally, we describe our experience and lessons learned with Muppet, which has been used extensively at Kosmix and WalmartLabs to power a broad range of applications in social media and e-commerce.

## 1. INTRODUCTION

MapReduce [8] has emerged as a popular paradigm to process big data. Using MapReduce, a developer simply writes a map function and a reduce function. The system automatically distributes the workload over a cluster of commodity machines, monitors the execution, and handles failures.

In the past few years, however, not just big data, but *fast data*, i.e., high-speed real-time and near-real-time data streams, has also exploded in volume and availability. Prime examples include sensor data streams, real-time stock market data, and social-media feeds such as Twitter, Facebook, YouTube, Foursquare, and Flickr. The emergence of social media in particular has greatly fueled the growth of fast data, with well over 4000 tweets per second (400 million tweets per day [12]), 3 billion Facebook likes and comments per day [9], and 5 million Foursquare checkins per day [2].

Numerous applications must process fast data, often with minimal latency and high scalability. For example, an application that monitors the Twitter Firehose for an ongoing earthquake may want to report relevant information within a few seconds of when a tweet appears, and must handle drastic spikes in the tweet volumes. As the number and sophistication of such applications grow, a natural question arises: *Can we provide a MapReduce-like framework for fast data, so that developers can quickly write and execute such applications on large clusters of machines, to achieve low latency and high scalability?*

In this paper we describe our investigation of this question, as carried out at Kosmix, a San-Francisco-Bay-Area startup, and at WalmartLabs, an advanced development lab newly established by Walmart (Walmart acquired Kosmix in May 2011 to form the seed of WalmartLabs). In Section 2 we describe a number of motivating applications that process fast data, and argue why MapReduce is not well suited for such applications.

In Section 3 we describe MapUpdate, a framework to process fast data. Like MapReduce, in MapUpdate the developer only has to write a few functions, specifically *map* and *update* ones. The system automatically executes these functions over a cluster of machines. MapUpdate, however, differs from MapReduce in several important aspects. First, MapUpdate operates on data streams, so map and update functions must be defined with respect to streams. For example, mappers and updaters map streams to streams, split streams, or merge streams. Second, streams may never end, so updaters use storage called *slates* to summarize the data they have seen so far. The notion of slates does not arise in MapReduce, nor in many recently proposed stream-processing systems (see Section 6). In MapUpdate, slates are in effect the "memories" of updaters, distributed across multiple map/update machines as well as persisted in a key-value store for later processing. Making such pieces of memory explicit and managing them as "first-class citizens," in a near-real-time fashion, is a key distinguishing aspect of the MapUpdate framework. Finally, a MapUpdate application often involves not just a mapper followed by an updater, but many of them in an elaborate workflow that consumes and generates data streams.

In Section 4 we describe Muppet, a MapUpdate implementation developed at Kosmix and WalmartLabs. We discuss the key challenges of Muppet in terms of distributed execution, managing slates, handling failures, reading slates, and sketch our solutions. Since mid-2010, we have used Muppet extensively to develop many social media and e-commerce applications (over streams such as Twitter and Foursquare). We describe this experience, lessons learned, as well as current and future extensions. We discuss related work in Section 6 and conclude in Section 7.

## 2. MOTIVATING APPLICATIONS

We describe several motivating applications, argue why MapReduce is not well suited for these applications, then outline our desiderata for the MapUpdate framework.

EXAMPLE 1. Consider an application that monitors the Foursquare-checkin stream to count the number of check-ins by retailer (e.g., JCPenney, Best Buy, and Walmart). For each incoming checkin, the application analyzes the text of the checkin (typically represented as a JSON object) to identify the retailer (if any), then increases the appropriate count. The counts are maintained continuously and displayed "live" on a Web page. □

EXAMPLE 2. The second application monitors the Twitter Firehose to detect hot topics as they occur. For ease of exposition, we will use the following simple heuristic: The application first classifies each incoming tweet into a small set of pre-defined topics. Next, as a pre-specified time interval (for example, a minute) passes, the application counts the number of tweets per topic. If this number divided by the average number of tweets that mention the same topic in the corresponding time interval each day (this average number is maintained by the application across multiple days) exceeds a pre-specified threshold, then the application emits the topic and the minute. Thus, the output is a stream of <topic, minute> pairs that reports which topic is hot for each minute. □

EXAMPLE 3. The third application maintains a reputation score for each Twitter user as users tweet. It analyzes each incoming tweet to determine if the tweet affects the score of any users, then changes those scores. The score of a user can be affected by many factors. For example, if a user $A$ retweets or replies to a user $B$, then the score of $B$ may change, depending on the score of $A$. The output is a real-time data structure (e.g., a hash table) of <user, score> pairs. □

Other applications include maintaining the top-ten URLs being passed around on Twitter, and maintaining live counters of the number of HTTP requests made to various parts of a Web site.

The key commonality underlying all of these applications is that they perform *stream computations*, which consume streams and produce streams or continuously-updated data structures as the output. We argue that MapReduce and variations of it are not well suited to such computations, for the following reasons. First, MapReduce runs on a static snapshot of a data set, while stream computations proceed over an evolving data stream. In MapReduce, the input data set does not (and cannot) change between the start of the computation and its finish, and no reducer's input is ready to run until all mappers have finished. In stream computations, the data is changing all the time; there is no such thing as working with a "snapshot" of a stream.

Second, every MapReduce computation has a "start" and a "finish." Stream computations never end. The data stream goes on forever. Typical stream computations update some data structure based on the input stream, and either output a stream or answer queries on the data structure they maintain (e.g., how many items have we seen so far that satisfy certain conditions?). In the MapReduce model, the reduce step needs to see a key and *all* the values associated with the key; this is impossible in a streaming model.

Finally, in case of failure, it is always possible (even if inconvenient) to restart a MapReduce computation from scratch. This possibility may not exist for many stream computations; streams continue to flow at their own rate, oblivious to processing issues. The system should be able to cope with failures very quickly to avoid falling too far behind the stream.

Thus, we need a new framework for stream computations. We would like this framework to satisfy the following requirements:

- The framework should be easy to program. It should have a simple model that enable the rapid development of many applications. Ideally, it should retain the familiar Map and Reduce feel, to help developers quickly write stream applications.

- The framework should manage dynamic data structures as first-class citizens. Many developers are accustomed to reasoning with such data structures explicitly in their code, and many stream applications need to produce such structures for higher-level applications.

- The framework should deliver low-latency processing. Applications should stay near real-time with their input streams, and computed data (i.e., dynamic data structures) should be available for live querying.

- The framework should scale up on commodity hardware with computation and stream rate.

In the next section we describe such a framework, MapUpdate.

## 3. THE MAPUPDATE FRAMEWORK

We assume a hardware platform similar to MapReduce, i.e., a cluster of commodity machines. In practice, the machines need to be more memory-heavy and less disk-heavy than in a MapReduce cluster. The reason is that most stream computations read streams as they flow by and maintain in-memory data structures, in contrast to MapReduce computations that read and produce large files.

**Events and Streams:** Example events are tweets, Facebook updates, and Foursquare checkins. Formally, an *event* is a tuple $\langle sid, ts, k, v \rangle$, where

- $sid$ is the ID of the stream that the event belongs to.

- $ts$ is a timestamp. To ensure a well-defined output when merging multiple streams, we assume timestamps are global across all streams (local timestamps, if any, can be stored in the value $v$).

- $k$ is a key. Keys have atomic values and need not be unique across events. For example, the key for a tweet might be the user ID. Key are used to group events, similar to the way they are used in MapReduce. We assume a global key space across all streams, though our model can be easily extended to handle multiple key spaces (e.g., one per stream).

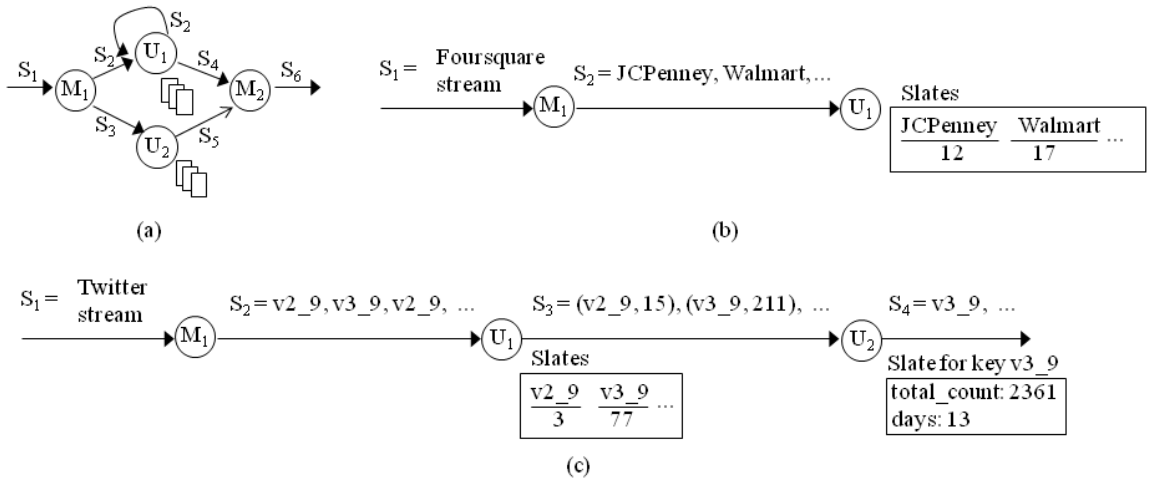

Figure 1: Example MapUpdate applications

- Finally, $v$ is a value, which can be any "blob" associated with the event. For example, if the event is a tweet with the key being the user ID, then the value can be the entire JSON object representing the tweet.

A *stream* is then a sequence of all events with the same *sid*, in the increasing order of timestamp *ts* (using a deterministic tie-breaking procedure). Streams can be external (e.g., the Twitter Firehose and the Foursquare-checkin stream) or internal, being generated by map and update functions as described below.

**Map Functions:** A map function $map(event) \rightarrow event*$ subscribes to one or more streams. All events from these streams will be fed as input, one by one, in the increasing order of their timestamps, using a deterministic tie-breaking procedure, to the map function. For example, suppose a map function $M$ subscribes to two streams $S_1$ and $S_2$, and suppose that $S_1$ begins with an event $e$ with timestamp 21:23 and that $S_2$ begins with an event $f$ with timestamp 21:25. Then event $e$ will be fed to $M$, followed by event $f$, followed by whichever event in $S_1$ and $S_2$ that has the next lowest timestamp, and so on.

Given an event as the input, the map function processes it then emits zero or more events to various streams. Thus, this function is analogous to the map function in MapReduce. Each output event has a timestamp greater than the timestamp of the input event, so that even if an output event is emitted to the same stream as the input event, the stream's events can be processed in timestamp order.

**Update Functions and Slates:** An update function $update(event, slate) \rightarrow event*$ also subscribes to one or more streams, and is also fed as input all events from these streams, one by one, in the increasing order of their timestamps, using a deterministic tie-breaking procedure. When the update function $U$ takes as input an event $e$ with key $k$, it is also given a slate $S_{U,k}$.

The slate $S_{U,k}$ is an in-memory data structure that stores all important information that the update function $U$ must keep about all the events with key $k$ that $U$ has seen so far. Within the update function $U$, there is one slate for each key. For example, if the key space is Twitter user IDs, then there is one slate per user ID for $U$, which may store summary information such as the number of tweets by that user, the time of the last tweet by the user, and the set of user interests that the update function has been able to infer from the tweets seen so far.

Recall that in MapReduce, the reduce function takes a key $k$ and the list $L$ of all values associated with key $k$, then "reduces" $L$ to emit new <key, value> pairs. Update plays an analogous role. However, because an update function operates on streams, it cannot take as input the list of *all* events with key $k$: It has not seen all such events, and in any case, the list of events with key $k$ that it has seen so far may already be too large to keep around.

To solve this problem, we introduce the notion of slate: a data structure that "summarizes" all events with key $k$ that an update function $U$ has seen so far. When given a new event $e$ with the same key, $U$ uses $e$ to update the slate (hence the name "update" for this function). Thus, the slate is a live data structure that is continuously updated in (near) real time. Each slate also has a time-to-live parameter, which is set to "forever" by default but can be set to a concrete value after which the slate can be deleted to free up memory. When an update function $U$ accesses a slate associated with a key $k$ for the first time (either because this is the first time $U$ sees an event with key $k$, or because the slate has been deleted after its time to live has expired), the update function must set up the set of variables it needs in the slate and initialize those variables.

It is important to emphasize that each pair <update $U$, key $k$> uniquely determines a slate, not that each key $k$ uniquely determines one. Indeed, we can have one slate associated with a key $k$ for an update function $U_1$, and yet another slate associated with key $k$ for another update function $U_2$.

In other words, unlike the "memoryless" map functions, an update function has a memory. This memory is partitioned into pieces called slates, each associated with a particular event key. Each copy of an update function, when run on a machine, is in charge of a set of event keys, and hence will directly update the set of slates associated with those keys (see Section 4.1). The slates are updated and kept in the main memory of those machines, but also persisted on disk in a key-value store (see Section 4.2). The developer can reason about the slates explicitly, and query them live.

In addition to updating the input slate, the update function may also emit new events, just like the map function.

**MapUpdate Applications:** A MapUpdate application is a workflow of map and update functions. Map functions consume streams and produce new streams, or emit new events into existing streams. Updater functions consume streams, continuously update slates, and produce new streams or emit new events into existing streams.

Thus, the workflow is modeled as a directed graph (allowing cycles), whose nodes represent map and update functions, and whose edges represent streams. Figure 1(a) shows such a graph. The output of the MapUpdate application is a set of streams and slates, as specified by the application. The following two examples illustrate MapUpdate applications:

EXAMPLE 4. Figure 1(b) shows the workflow of the application that counts Foursquare checkins per retailer (see Example 1). This application starts with stream $S_1$, the Foursquare checkin stream. A map function $M_1$ inspects each checkin to see if the checkin happened at a retailer's location. If yes, $M_1$ emits an event with the retailer ID (e.g., JCPenney, Walmart) to a new stream $S_2$. An update function $U_1$ subscribes to $S_2$ and counts the number of checkins per retailer. Specifically, for each retailer $U_1$ maintains a slate with a *count* variable initially set to 0. $U_1$ then increments *count* by 1 every time it sees an event with the same retailer ID. The output of the application is the set of slates maintained by $U_1$. □

EXAMPLE 5. Figure 1(c) describes the application that monitors tweets to detect hot topics (see Example 2). The application starts with $S_1$, the Twitter stream. A map function $M_1$ analyzes each tweet $t$ in $S_1$ to infer a set of topics (taken from a predefined set of possible topics). Let $m$ be the minute in which the timestamp of tweet $t$ occurs (e.g., if the timestamp is 00:14 then $m = 14$; if the timestamp is 23:59 then $m = 1439$). For each inferred topic $v$, $M_1$ publishes an event with the key $v\_m$ (i.e., a string that concatenates $v$ and $m$) to a new stream $S_2$ indicating that topic $v$ is mentioned in a tweet that occurs in the minute $m$.

An update function $U_1$ subscribes to stream $S_2$ to count the frequencies of topics per minute. When $U_1$ first encounters an event with key $v\_m$, it creates a slate for this key, and sets *count* $= 0$ in the slate. Every time it sees an event with the same key, it increments *count* by 1. After a minute (counting from when it sees the first event with key $v\_m$), $U_1$ publishes an event (key $= v\_m$, value $=$ *count*) to a new stream $S_3$, indicating that topic $v$ is mentioned *count* times in the minute $m$.

An update $U_2$ monitors stream $S_3$ to find hot topics. Specifically, when $U_2$ sees an event ($v\_m$, *count*), it computes *count*$/avg\_count_{v\_m}$. If this ratio exceeds a certain threshold then $U_2$ publishes an event with key $v\_m$ to a new stream $S_4$, indicating that topic $v$ is hot in the minute $m$. The quantity $avg\_count_{v\_m}$ is the average count of topic $v$ in the minute $m$. $U_2$ keeps a slate for key $v\_m$, with two summaries:

- *total_count*: the total number of times topic $v$ has been mentioned in the minute $m$ so far, since the first day, and

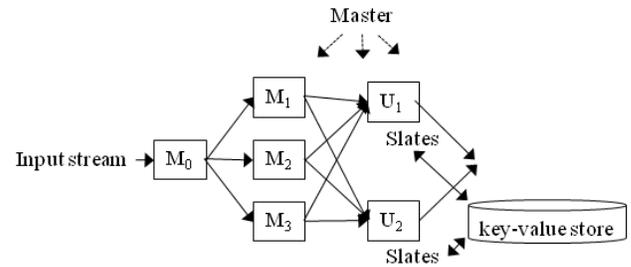

**Figure 2: An illustration of the working of Muppet**

- *days*: the number of days since when the application is deployed.

$U_2$ uses these two quantities to compute $avg\_count_{v\_m}$. The output of the application is stream $S_4$. □

We note that update functions maintain no "global" variables (across the slates). For example, update function $U_2$ in Example 5 maintains a "local" variable *days* in *each* of the slates, even though all *days* variables have the same value across all slates, and thus technically can be "merged" into a single "global" variable. Update functions do not maintain global variables across the slates to avoid the concurrency-control problem in which multiple copies of the same update functions, run on different machines, all attempt to modify the same "global" variables at the same time.

As described above, it is not difficult to show that if

- the map and update functions are deterministic, in that the input event completely determines the output events and slate updates;

- the events of the subscribed streams are fed into a map/update function in a well-defined order (which in this case is the increasing order of their timestamps, using a deterministic tie-breaking procedure); and

- the timestamps of output events are greater than that of the input event (to ensure that loop executions are well-defined),

then a MapUpdate application is well-defined, in that it generates well-defined streams and sequences of slate updates. Ideally, a MapUpdate implementation should produce these exact streams and slate updates. Due to practical constraints, however, it often can only approximate them, but should try to do so as closely as possible.

To write a MapUpdate application, a developer writes the necessary map and update functions, then a configuration file that includes the workflow graph.

## 4. THE MUPPET SYSTEM

We now describe Muppet, our implementation of the MapUpdate framework. We describe Muppet 1.0 (Sections 4.1–4.4), developed at Kosmix, then Muppet 2.0 (Section 4.5), developed at WalmartLabs, which addresses several key limitations of Muppet 1.0.

## 4.1 Distributed Execution

To execute a MapUpdate application on a cluster of machines, Muppet starts up a set of programs on each machine. Each program executes a map or update function. The programs are called *workers*, which can be divided into *mappers* and *updaters*, depending on which function they run. To distribute the computation, each worker will be fed only events of certain key values, as determined by a hash function. Figure 2 illustrates this process. Given an application that runs a map function followed by an update function, suppose Muppet has decided to run five workers: three mappers $M_1-M_3$ for the map function, and two updaters $U_1-U_2$ for the update function.

Muppet begins by using a special mapper $M_0$ to read from the input stream (see the figure). Given an event $e$ with a key $k$, $M_0$ hashes $k$ to find out which mapper to send $e$ to. Suppose this mapper is $M_1$. $M_0$ places event $e$ in the queue for $M_1$ (this queue is maintained in memory for the $M_1$ program). $M_0$ then reads the next event in the input stream, and so on.

The mapper $M_1$ takes the next event from its queue, processes the event, and produces a set of events. For each produced event $f$, $M_1$ hashes its key and the destination update function to find out which updater to send the event to. Suppose this updater is $U_2$. $M_1$ then places event $f$ in the queue for $U_2$. $M_1$ then processes the next event from its own queue, and so on.

Thus, events flow continuously through the workflow of mappers and updaters. For each event that a worker produces, it must find out which workers to send the event to, where those workers are (i.e., on which machines), then place the event into the queues for the workers. One way to make this determination (and the way Muppet currently employs) is to give all workers the same hash function to map <event key, destination map/update function> to workers. That way, after producing an event, any worker can instantly calculate which worker the event hashes to, then contact that worker to place the event into the appropriate queue. Each worker has its own queue for input events.

This mode of passing events is in stark contrast to MapReduce. There, after a mapper has produced output files, it contacts the master, which in turn notifies the appropriate reducers to get the files. This solution is not well suited for our setting because it is very important for many MapUpdate applications to minimize latency, i.e., to produce streaming output data quickly, in as "real time" as possible. To achieve this, we try to remove as many intermediaries as possible. Hence, Muppet lets the workers pass events directly to one another without going through any master. (The master in Muppet is used for handling failures, see Section 4.3.)

Hashing ensures that all events with the same key $k$ will go to the same updater $U$ (this is similar to MapReduce, where all events with the same key go to the same reducer). The updater uses the events to update a slate $S_{U,k}$ associated with key $k$. Only this updater can update $S_{U,k}$, so there are no concurrent updates for $S_{U,k}$.

## 4.2 Managing Slates

**Persisting Slates in a Key-Value Store:** As described, an updater $U$ maintains a slate $S_{U,k}$ for each key $k$. These slates are cached in the memory of the machine running $U$. Muppet also persists them in a key-value store, for three reasons. First, the slates of $U$ may outgrow the memory, in which case some of them have to be spilled to disk. Second, persistent slates help resuming, restarting, or recovering the application from crashes. Finally, we often need to query the slates, which represent the computation of a MapUpdate application, long after the termination of the application.

Muppet currently uses Cassandra as the key-value store. A Cassandra cluster consists of a set of machines, each running the Cassandra program, all configured to recognize one another as parts of the same cluster. The cluster maintains a set of key spaces, each of which contains a set of column families. Each column family, in turn, stores data values indexed by <key, column> pairs.

A Muppet application's configuration file identifies a Cassandra cluster (by its machine names and service TCP port), a key space within the cluster, and a column family within the key space. Within this column family, Muppet stores slate $S_{U,k}$ (for the update function $U$ and key $k$) as a value at row $k$ and column $U$. Our applications often use JSON to encode slates for language independence and flexibility, so Muppet compresses each slate before storing it in the key-value store.

When the updater $U$ needs the slate with key $k$, Muppet first checks the cache (in the memory of the machine running $U$). If the slate is not found, Muppet retrieves the slate from the Cassandra cluster by reading the value indexed by the pair $<k,U>$. The retrieved value is decompressed then passed to the updater.

If the requested slate does not exist in Cassandra, either because the updater has never seen an event with this key, or because the slate has been deleted after its time to live expired, then Muppet initializes a new slate in the cache, then passes it to the requesting updater.

**Using SSDs and Caching Slates:** We run our Cassandra key-value store on solid-state flash-memory storage (SSDs). This allows us to devote Cassandra's memory to buffering writes, while caching reads in the slate cache (i.e., the memory of machines running updaters). We found this solution very helpful for several reasons:

- When Muppet starts up, its slate cache is empty, so early update events may require many row fetches from the key-value store. Fast random access helps the store respond to this volume of reads more quickly, warming the slate cache.

- While running, Muppet often needs random-seek I/O capacity to fetch uncached slates. Meanwhile, Cassandra also requires I/O capacity for periodic compactions, thus slowing down Muppet. Using SSDs provides high I/O capacity to help us sustain both needs.

- Because applications often update popular slates repeatedly, we minimize disk I/O for writing at the key-value store if we devote the store's main memory to buffering writes. Overwrites of the same row in the key-value store are relatively inexpensive if the row is still in memory at the time of the write, so it is advantageous for us to delay flushing the writes (i.e., the memory table) to disk as long as possible. Further, the more times a row is flushed to disk by the store since its last file compaction, the more files will have to be checked for the row when it needs to be retrieved.

**Flushing, Quorum, and Time-to-Live Parameters:**
Muppet applications can adjust a set of parameters to reach the desired level of performance, reliability, and consistency. First, dirty (updated) slates are periodically flushed to the key-value store. The application can set the flushing interval, ranging from "immediate write-through" to "only when evicted from cache."

Second, the application can specify the desired quorum used by the Cassandra store for a successful read/write operation: any single machine to which the data is assigned for storage, a majority of replicas where the data is assigned, or all of the replicas where the data is assigned.

Third, key-value stores like Cassandra allow their clients to specify a time-to-live (TTL) parameter for each write. Correspondingly, each updater function in a Muppet application can have a TTL value configured for its slates. Slates that have not been updated (written) for longer than the TTL value may be garbage-collected by the key-value store, resetting to an empty slate at that time.

The TTL parameter helps contain the amount of storage used by a Muppet application over time. Many such applications only care about current activities in their streams, declining to receive or generate events on obsolete keys. For example, an application may want to keep track of only active Twitter users (e.g., those who have tweeted at least once in the past quarter), a working set which is typically much smaller than the set of all Twitter users who have ever tweeted.

By making TTL a user-configured parameter, application developers can keep slates as long as needed without having to manually delete slates that are no longer useful. This setting is configurable per update function because different update functions often track different kinds of data, thus requiring different shelf lives.

### 4.3 Handling Failures

We now describe how Muppet handles two major types of failure: machine crash and queue overflow.

**Machine Crash:** In Muppet each worker keeps track of all failed machines. Recall that when a worker $A$ needs to pass an event, it determines the worker $B$ to which to send the event by hashing the key and destination updater function of the event (technically accomplished using a hash ring). Worker $A$ reaches worker $B$ to place the event on $B$'s incoming-event queue.

If $A$ cannot contact $B$, then it assumes the machine hosting $B$ has failed, and $A$ contacts the master to report the machine failure. The master broadcasts the machine failure to all workers, which update their lists of failed machines accordingly. Since all workers use the same hash ring, from then on all events with the same key will be routed to worker $C$ instead of the (now failed) worker $B$. The event that failed to reach $B$ is lost (and logged as lost) rather than sent through the event-dispatch process again.

In Muppet, since events typically flow through the system at high speed, and since a worker is frequently contacted, in most cases the above solution allows us to detect worker failures and recover from them in a timely fashion, and is preferable to the MapReduce solution of having the master pinging the workers periodically to detect worker failure.

When an updater fails, whatever changes that it has made to the slates and that have not yet been flushed to the key-value store are lost. Furthermore, all events in its queue are also lost. Currently, low latency is far more important for most of our Muppet applications, while failing to process some tweets, for example, is acceptable. Hence, we do not attempt to recover the lost events in the queue. Instead, we focus on quickly detecting the failed worker and redirecting events to another worker, thereby minimizing our latency and losses. Developing a replay capability to recover the lost events is a subject of future work.

**Queue Overflow:** When a worker $A$ tries to place an event into the queue of a worker $B$, if the queue of $B$ is full (i.e., its size has reached a pre-specified limit), $B$ will decline to accept the event. In this case $A$ has to invoke a queue overflow mechanism.

The queue overflow mechanism can take one of several actions. First, it can decide to drop the incoming events (until $B$ can accept events again). The dropped events can be logged for later processing and debugging. Second, it can direct the incoming events to a specified "overflow" stream whose recipients can process such events. The overflow stream can be connected to map and update functions that implement "slightly degraded" service, for example by substituting expensive operations in the main workflow pipeline with approximate operations that are cheaper to execute. Finally, the overflow mechanism can also decide to slow down the pace of passing events among the mappers and updaters (as discussed in more detail in Section 5).

### 4.4 Reading Slates

As Muppet runs a MapUpdate application, the application maintains live state in its updaters' slates. This state is often the value of the application's computation, and is often read by higher-level applications. To make this possible, Muppet provides a small HTTP server on each node for slate fetches.

The URI of a slate fetch includes the name of the updater and the key of the slate to uniquely identify a slate. The fetch retrieves the slate from Muppet's slate cache (on the appropriate machine, forwarding the request internally if necessary) rather than from the durable key-value store to ensure an up-to-date reply.

### 4.5 Developing Muppet 2.0

So far we have described Muppet 1.0, which was developed at Kosmix. In Muppet 1.0, each worker was implemented as two tightly coupled processes: a Perl process called a conductor, and a process running the JVM called a task processor here.

The conductor is in charge of all "Muppet logistics," including retrieving the next event from its queue of incoming events; sending the event (together with a slate, if necessary) to the JVM task processor; receiving the output events (and a modified slate if applicable) from the JVM task processor; hashing the output events to their appropriate destinations; enqueueing the events at their destination workers' queues, and so on. The JVM task processor's sole task is to run the map or update code to process the event passed to it by the conductor, then send the output events back to the conductor.

As described, Muppet 1.0 suffered from several limitations:

- Recall that a machine typically runs a set of workers. Each worker on the machine must load its own copy

of the map or update code so that it can run its JVM task processor. These duplicate copies of code waste memory.

- Passing data between processes (e.g., passing events back and forth between the conductor and the task processor) can be computationally wasteful.

- Each worker on a machine maintains its own slate (in the conductor). Thus, the slate cache on the machine is technically the set of disparate slates maintained by the workers. Maintaining the slates disparately can result in a significant waste of memory. For example, suppose we determine that we need to cache a working set of 100 popular slates on a single machine to run update events efficiently. If we run a single updater on the machine, we could reasonably assign the update function a slate cache of 100 slates to capture this working set. By contrast, if we run five updaters on the same machine, Muppet divides the slates of the update function among them. Because the keys of the popular slates may be hashed unevenly among them (for example, one of the five updaters might get 25 of the popular slates, not 20), we have to configure a larger slate cache per updater (e.g., 25 slates each and not 20) to cache the same working set (yielding a larger total slate cache of 125 slates instead of 100).

- Finally, it is difficult to fully utilize the number of cores on the machine, because the number of workers per machine is typically set based on the nature of the application, not based on the number of cores. In a machine with numerous CPU cores, it may be impractical to run as many workers as cores for every map and reduce function to utilize all cores regardless of which function has the most events to process at any moment. As the number of cores and the number of map and reduce functions grow, the number of workers would grow, amplifying the memory problems described above. The more numerous processes can also require more context switching to execute when events distribute widely among them.

Muppet 2.0, developed at WalmartLabs mostly in Java and Scala, addresses these limitations. In Muppet 2.0, we start up many threads of execution in a dedicated thread pool per machine. Each thread in this thread pool is now a *worker*, capable of running *any* map or update function. It is helpful to specify as large a number of threads as the parallelization of the application code allows. For example, the number may be as large as the number of CPU cores available on a machine, or smaller if the application's operators depend on a bottleneck resource that has a lower parallel-scaling limit.

Besides the worker threads, each machine also runs a thread to provide background I/O to the durable key-value store (so that writes to the store can proceed without blocking map and update calls), and a thread pool to provide HTTP service for slate reads and basic status information (such as the event count of the largest event queues).

Map and update functions are then written so that they can be run in multiple threads concurrently. To conserve memory, each map and update function is constructed only once and shared by all threads. All slates are now kept in a single "central" slate cache, not scattered in multiple conductor processes as in Muppet 1.0.

To process events, Muppet 2.0 maintains a queue per worker thread. When an event arrives at the machine, it is hashed by event key and destination updater function into a primary event queue and a secondary event queue. If the thread for either queue is already processing this event key for this update function, then the event is placed in the corresponding queue. Otherwise, the event is placed in the primary queue unless the secondary queue is significantly shorter, in which case the event is placed in the secondary queue instead. Each thread then takes the next event from its queue; executes the map or update function, depending on what the event requires; updates the appropriate slates if necessary; sends out the output events; takes the next event from its queue; and so on.

The dispatch of an incoming event to only one of two target queues, instead of to potentially any of the queues, brings several benefits. First, an incoming event locks no more than two queues to be dispatched to one of them regardless of the number of threads running map and update operations, reducing queue-lock contention when receiving incoming events.

Second, events of the same key for the same update function do not scatter across many threads on the same machine, reducing contention for the same slate among threads when those events get executed.

Finally, should an incoming event's primary queue be already heavily loaded by some other events, the incoming event can be placed on a secondary queue to better balance event load across available cores.

Thus, unlike Muppet 1.0, in which only one worker can process events of the same key for a particular update function, ensuring no slate contention, in Muppet 2.0, two workers can vie for the same slate, but this contention is limited to at most two workers per slate.

A fundamental reason why Muppet 2.0 allows slate contention is that if only one worker can process events of the same key, that worker can become a hotspot: if it is overloaded by a huge number of events with key $k_1$ already in its queue, a long time may pass before the worker gets around to processing events with some key $k_2$. Hence, Muppet 2.0 allows events with key $k_2$ to be placed into the queue of a second worker, if the queue of the first worker is already too long. This helps relieve the hotspot at the first worker, but can introduce slate contention when both the first worker and the second worker get events with key $k_2$ enqueued. In practice we have found that if the contention for any slate is limited to just two workers, it does not cause noticeable problems for our current applications.

As described, Muppet 2.0 addresses the above four limitations of Muppet 1.0. Each worker is now a thread that can execute any map or update function, not a pair of tightly coupled processes that can execute a single map or update function. All threads share the same map and update code, thus eliminating the waste of memory to hold redundant copies of the code. Passing data between processes is eliminated within each machine. All slates are now kept in a central pool, eliminating potentially underutilized slate-cache memory. Finally, the number of worker threads are set to maximize the potential for parallel execution on multicore machines.

# 5. MUPPET EXPERIENCE AND ONGOING EXTENSIONS

The first version of Muppet went into production at Kosmix in mid-2010. Since then we have improved Muppet several times, as discussed above, and used it extensively at Kosmix and later at WalmartLabs. At Kosmix it was used to process the Twitter Firehose and Foursquare-checkin stream. By early 2011 Muppet processed over 100 millions tweets and 1.5 million checkins per day. It kept over 30 millions slates of user profiles and 4 million slates of venue profiles. It ran over a cluster of tens of machines, and achieved a latency of under 2 seconds. Muppet was used to power TweetBeat, the flagship product of Kosmix, and now ShopyCat, a popular Facebook product recently released by WalmartLabs. About 16 developers (about half of the developer workforce at Kosmix) have used Muppet to quickly write about 15 applications, a number of whom have worked with Muppet applications at Kosmix and selected it again for new applications at WalmartLabs. By June 2012, our Cassandra store has grown to maintain over 2 billion slates for various production Muppet applications. We now discuss our experience running Muppet and several ongoing extensions.

**Limiting Slate Sizes:** We observe that slates can grow quite large and updaters that maintain large slates can run more slowly due to the overhead. Consequently, we encourage developers to keep individual slates small, e.g., many kilobytes rather than many megabytes.

**Changing the Number of Machines on the Fly:** Muppet runs on a cluster of machines. Currently the number of machines in the cluster cannot be changed on the fly. To add more machines, for example, we have to restart the Muppet application. While this setting has proven sufficient for our applications so far, one can imagine scenarios where it is desirable to be able to change the number of machines on the fly.

Hence, we are currently exploring this option. The main challenge is how to redistribute the workload. For example, suppose that a machine $A$ is currently overwhelmed with processing events with key $k$. So we want to add a new machine $B$ to help with this. Should we move some of the events with key $k$ to machine $B$? If so, both machines $A$ and $B$ will be processing events with key $k$. The slate for these events would have to be replicated at both $A$ and $B$; and coordinating the two slate copies will be highly difficult.

**Handling Hotspots:** The distribution of event keys can be strongly skewed (e.g., follow a Zipfian distribution). Consequently, updaters can receive widely varying loads, and an updater that receives an overwhelming load can potentially become a hotspot.

We already discussed one way to handle such hot spots: sending events with the same key to up to two threads instead of one (see Section 4.5). This approach allows Muppet to make progress on processing events with the same key on a secondary thread if the first thread is currently bogged down with other events, and at the same time reduce the workload of the first thread. This load distribution comes at the cost of some contention between the two threads for ownership of the same slate.

Another way to handle hotspots is to exploit the fact that numerous update computations are associative and commutative, to distribute the workload of an overwhelmed updater among a set of updaters. The following simple example illustrates this idea:

EXAMPLE 6. Consider again the application that counts Foursquare checkins per retailer in Example 4. In this application, a map function examines each checkin to emit the name of a retailer (if any), such as JCPenney, Best Buy, or Walmart. An update function then counts the emitted location events per retailer.

Let $U$ be the updater that counts Best Buy events. Suppose, hypothetically, that a lot of people are checking into Best Buy: $U$ can quickly become a hotspot as it becomes overwhelmed by the number of arriving Best Buy events. To address this problem, observe that counting Best Buy events is associative and commutative. Hence, instead of using just a single updater $U$, we can use a set of updaters, each of which counts just a subset of Best Buy events. We can then sum the counts of these updaters.

Specifically, we can modify the map function to replace the single key "Best Buy" with two keys "Best Buy1" and "Best Buy2," say. In effect, the map function partitions the set of events with key "Best Buy" into two subsets with keys "Best Buy1" and "Best Buy2," respectively. Next, we modify the update function so that it regularly emits the counts of "Best Buy1" events and "Best Buy2" events, respectively, as new events under the key "Best Buy." Finally, we write a new update function that receives the events of key "Best Buy" to determine the total counts of "Best Buy1" events and "Best Buy2" events. □

We have discussed how to redistribute the workload of a hotspot updater among a set of updaters. Yet another way to handle hotspots is to slow down the pace of events in the workflow. In our settings, some Muppet applications do not need near-real-time latencies (e.g., in milliseconds or seconds). Examples include applications that do not use time-related or time-sensitive data (and simply tap the MapUpdate framework as a convenient way to implement a workflow for machine-scalable distributed computation) and applications that run on legacy tweets. In such cases, accepting longer latencies for stable operation is often acceptable. Consequently, when Muppet detects a hotspot, it can slow down the pace at which it consumes events from its *input streams* (e.g., the Twitter Firehose) to allow until the hotspot updater has a chance to catch up. We call this approach *source throttling*.

It is also possible to throttle the pace of events at *any later point* in the workflow, not just at the input streams, but if not done very carefully, doing so can quickly introduce deadlocks. To see why, consider an updater $U$ that emits events into its input streams (thus introducing a loop). Suppose $U$ consumes an event $e$ and is about to emit 10,000 events back into $U$'s input stream, and emitting 10,000 events all at once would overwhelm $U$. We may be tempted to slow down the pace at which events are emitted: Muppet could emit 10,000 events one by one, in an incremental fashion, as soon as $U$ is ready to consume its next event. Unfortunately, this approach would introduce a deadlock. After emitting the first event of the 10,000 output events, Muppet would be waiting for $U$ to finish processing the current event (i.e., event $e$), before emitting the second event of the 10,000 output events. However, $U$ cannot finish processing event $e$ until we have emitted all 10,000 output events.

Note that the above scenario does not arise in the case of slowing down the pace of consuming events from the application's input streams (e.g., the Twitter Firehose), because we assume that no mappers nor updaters can emit events into such streams.

**Placing Mappers and Updaters:** Currently the placement of mappers and updaters in Muppet is in effect decided by the hashing function that hashes event keys to machines and workers. We are exploring how to place mappers and updaters so that they are close to their data in a way that reduces network traffic.

This problem is nontrivial in part because Muppet may not know in advance which event streams will have the most data. For example, let us revisit the application that counts Foursquare checkins per retailer in Example 4. In it, a map function emits an event to an update function each time a checkin for a recognized retailer arrives. Suppose, for simplicity, that checkins arrive at a particular machine $m$, and a mapper there runs the map function. Which keys (and corresponding slates) for the update function should go to machine $m$, and which ones should be assigned elsewhere? If the most popular retailers' slates reside on machine $m$, then the smallest number of events from the mapper have to traverse the network (and pay the corresponding latency costs or consume the corresponding network utilization) to reach an updater. Unfortunately, such a determination depends on the contents of the checkin events themselves, so Muppet cannot determine this assignment in advance. Muppet cannot even know whether perturbations in retailer popularity are transient spikes to absorb or changing trends that require a different slate-to-machine assignment. Finally, applications typically have multiple update functions that may be directly or indirectly connected by event flows, so moving slates to optimize network traffic into one update function may affect the network usage of events from it to a subsequent update function. (Key and slate assignments that reduce network traffic for the input or output of one functions may increase the network traffic coming in or out another function.)

**Bulk Reading of Slates:** In many applications at Kosmix and WalmartLabs users want to make periodic dumps of many slates. In such cases, repeated HTTP slate fetches can be expensive (in network round trips) or difficult (because the query agent must know all the slate keys in advance to enumerate the slate requests).

To address this problem, we have advised bulk-dump users to log the relevant slate data that they wish to process in bulk later as a part of the applications' update functions. This approach allows users to write less than the entire slate to minimize the dumped data, and provides steady-state write behavior that avoids sudden bulk I/O, which can affect the performance of the machines supporting the application. These writes can be streamed using a library of the user's choice into HDFS, for example, if further processing in Hadoop is desired.

Another approach users can take is to request large-volume row reads from the durable key-value store itself. Users that choose to do so must know how slates are written to the key-value store (an implementation detail of Muppet, described in Section 4.2) to extract the slates back from the appropriate key-value-store queries.

In the future, we would like to revisit the scenario of how users use MapUpdate-application slates for later Hadoop processing so that we can simplify and automate this integrated use case better.

**Managing Side Effects:** We have found that applications sometimes wish to act on events in ways outside of updating slates or publishing events. Such actions fall outside the scope of the current MapUpdate framework itself, and we currently leave it to the application to carry out such actions in its own map and update functions. For example, applications may want to log relevant slate data for later bulk processing, as discussed above. As another example, developers often instrument map and update functions to log certain data for later debugging. As yet another example, an application may wish to have a map or update function make some HTTP request to a server when a criterion is satisfied, so that the outside server can be notified when the criterion is satisfied rather than requiring the server to sample or probe slates repeatedly to make the determination.

While leaving it to the applications to carry out such side-effect actions, we do advise developers to be careful of subtle effects of such actions on the Muppet application. For example, asking mappers and updaters to write to a common log can introduce lock contention for the common logger, thereby dramatically slowing down the workers.

## 6. RELATED WORK

We have compared our work with MapReduce throughout this paper.

A number of recent works (such as MapReduce Online [7], Nova [18], work by Li et al. [14], and Incoop [4]) have extended MapReduce to perform incremental batch processing. MapReduce Online pipelines data between the map and reduce operators by calling reduce with partial data for early results. To retain the MapReduce programming model, it runs reduce periodically (as a minimum interval of time passes or a batch of new data arrives), retaining some of its blocking behavior. Nova determines and provides the deltas between increments directly to workflows written for Pig, but its authors warn that this approach is more suitable for large batches than small increments because of the overhead costs in underlying systems. Systems such as Incoop apply memoization to the results of partial computations so that subsequent computations can efficiently reuse results for inputs that were unchanged by additional incremental data. The prototype one-pass analytics platform described in [14] optimizes MapReduce jobs by (among other improvements) exploiting main memory to pre-combine map outputs by key (when the MapReduce job has an optional combine function defined); this optimization is most nearly analogous to how Muppet exploits main memory to cache slates (indexed by updater operator and event key), minus any event-serialization considerations for a slate.

By contrast, MapUpdate uses slates to *summarize* past data, so an updater can immediately process each event (and change the slate) as the event comes in. This approach allows us to stream events through the system with millisecond to second latencies.

Many streaming query-processing systems, such as continuous query systems, have been developed in the database community [10] and in industry (e.g., Aurora [21], commercialized as StreamBase Systems, and Borealis [1]; Cloud-Scale [6]; STREAM [3]; SPADE [11] for System S, com-mercialized as IBM InfoSphere Streams; and Telegraph [5], commercialized into Truviso). Our work differs from these systems in two important aspects. First, these systems often employ declarative query languages over structured data with known schema. In contrast, we make few assumptions about the structure of the data, and adopt a MapReduce style in which applications are decomposed into a procedural workflow of custom code. Second, much work has focused on optimizing query processing over data streams in a relational-database style (e.g., how to factor operations out of multiple queries, and push operators to optimal locations for query execution). In contrast, we focus on how to efficiently execute relatively arbitrary Maps and Updates over a cluster of machines to achieve low latency and high scalability. Like Flux [19], Muppet strives to distribute the input load of each update operator across multiple machines, but Muppet does not currently change its load partitioning dynamically except when a machine fails. Unlike Spark Streaming [20], which offers APIs for Scala to enable developers to write programs modeling streams as a sequence of small batch computations, Muppet offers a simple MapReduce-style framework, MapUpdate, to enable developers to write continuously updating streaming applications.

Other avenues of low-latency-application development are available, including specialized stream-processing chips such as GPUs programmed with OpenCL [13] or CUDA [17]), and high-bandwidth remote-memory access (RDMA) over specialized high-speed interconnects (such as InfiniBand). Unlike computations on GPUs, which often perform well computing similar operations in parallel on a vector of values, MapUpdate is designed to allow arbitrary computation on general-purpose CPU cores for each input event, including data-dependent recursion or event publication. Unlike RDMA, which allows an application to span multiple machines using a shared-memory model implemented on high-speed networks, Muppet is designed to run on commodity hardware, allowing us to build large slate caches using the union of main memory on multiple machines linked by inexpensive gigabit Ethernet networks. The MapUpdate model, in particular, allows us to explicitly shard application state across machines to sidestep an explicit need for fast shared memory between them.

Our work is also similar in spirit to recently proposed distributed stream-processing systems, such as S4 [16] and Storm [15]. These systems, however, leave it to the application to implement and manage its own state. Our experience suggests that this is highly nontrivial in many cases. By contrast, Muppet transparently manages application storage, which are slates in our case, and makes these slates accessible as the continually-updated computed values of a streaming application.

Indeed, the explicit and first-class-citizen management of application memory in the form of slates is a key distinguishing aspect of our work, in sharp contrast to current work in incremental MapReduce, RDBMS-style stream processing, and industrial distributed stream processing systems.

## 7. CONCLUSION

We have motivated the need for a MapReduce-style framework for processing fast data. We have described such a framework, MapUpdate, and our implementation of MapUpdate at Kosmix and WalmartLabs, Muppet. Throughout the discussion we have tried to motivate and highlight the differences between MapReduce and MapUpdate. The key differences include the need to redefine applications and user-defined functions to operate over streams; the need for slates, continuously updated data structures that "summarize" the events seen so far; the need for workflows of Maps and Updates; the importance of minimizing latency and how that influences design decisions on distributing and passing events and handling failures; and finally the need to persist slates, as the semantics of the application dictates.

We have also reported on our experience using Muppet at Kosmix and WalmartLabs. Overall, we conclude that a MapReduce-style framework to process fast data, such as implemented in Muppet, is feasible and highly promising, in terms of allowing developers to quickly write fast-data applications, and to achieve low latency and high scalability in those applications. Learning from our experience, we are currently deploying Muppet to more applications and are extending it in several important directions.

# APPENDIX
## A. EXAMPLE MAP AND UPDATE

The map and update functions are expressed in JVM languages for Muppet as application-provided class implementations of Java interfaces called Mapper and Updater. Implementations of the interfaces are constructed using two parameters, a configuration object for the application and a string for the name of the map or update function being instantiated. (Because the same Mapper or Updater code can be reused as different map and update functions throughout an application, each map and update function in the application is identified by unique name.)

Figures 3 and 4 (cf. $M_1$ and $U_1$ in Example 4, respectively) show an example of how these interfaces could be used in Java.

```java
package com.walmartlabs.example;

import java.nio.charset.Charset;
import java.util.regex.Matcher;
import java.util.regex.Pattern;

import org.slf4j.Logger;
import org.slf4j.LoggerFactory;

import com.kosmix.muppet.application.Config;
import com.kosmix.muppet.application.binary.Mapper;
import com.kosmix.muppet.application.binary.PerformerUtilities;

public class RetailerMapper implements Mapper {
    private final Logger logger =
        LoggerFactory.getLogger(RetailerMapper.class);
    private final Charset charset = Charset.forName("UTF-8");

    private final Pattern walmart =
        Pattern.compile("(?i)\\s*wal.*mart.*");
    private final Pattern samsclub =
        Pattern.compile("(?i)\\s*sam.*s\\s*club\\s*");

    private String name;

    public RetailerMapper(Config config, String n) { name = n; }

    @Override
    public String getName() { return name; }

    @Override
    public void map(PerformerUtilities submitter,
                String stream, byte[] key, byte[] event)
    {
        String checkin = new String(event, charset);
        String venue = getVenue(checkin);

        String retailer = null;
        if (walmart.matcher(venue).matches()) {
            retailer = "Walmart";
        } else if (samsclub.matcher(venue).matches()) {
            retailer = "Sam's Club";
        }

        if (retailer != null) {
            try {
                submitter.publish("S_2",
                    retailer.getBytes(charset), event);
            } catch(Exception e) {
                logger.error("Could not publish event: "+
                    e.toString());
            }
        }
    }

    private String getVenue(String checkin) {
        // actual checkin parsing would go here
        return "name of venue";
    }
}
```

**Figure 3: An example Java-language Mapper**

```
package com.walmartlabs.example;

import java.nio.charset.Charset;

import org.slf4j.Logger;
import org.slf4j.LoggerFactory;

import com.kosmix.muppet.application.Config;
import com.kosmix.muppet.application.binary.PerformerUtilities;
import com.kosmix.muppet.application.binary.Updater;

public class Counter implements Updater {
    private final Logger logger =
        LoggerFactory.getLogger(Counter.class);
    private final Charset charset = Charset.forName("UTF-8");

    private String name;

    public Counter(Config config, String n) { name = n; }

    @Override
    public String getName() { return name; }

    @Override
    public void update(PerformerUtilities submitter,
                       String stream, byte[] key, byte[] event,
                       byte[] slate)
    {
        int count = 0;
        try {
            if (slate != null)
                count =
                    Integer.parseInt(new String(slate, charset));
        } catch (NumberFormatException e) {
            count = 0;
        }
        ++count;
        byte[] updatedSlate =
            Integer.toString(count).getBytes(charset);
        submitter.replaceSlate(updatedSlate);
    }
}
```

Figure 4: An example Java-language Updater